\def\rfr#1{eq. (\ref{#1})}

\def\derp#1#2{\rp{\partial{#1}}{\partial{#2}}}

\def\virg#1{``#1''}

\def\eqi{\begin{equation}}
\def\eqf{\end{equation}}
\def\eqia{\begin{eqnarray}}
\def\eqfa{\end{eqnarray}}
\def\rp#1#2{{#1\over#2}} \def\lb#1{\label{#1}}

\documentclass{aastex}
\usepackage{url}
\usepackage{amsmath,amsthm,amscd,amssymb}
\usepackage{latexsym}
\usepackage{graphicx,epsfig}

\RequirePackage{color}

\newcommand{\emaila}{lorenzo.iorio@libero.it}

\begin{document}

\title{Phenomenological constraints on accretion of non-annihilating dark matter on the PSR B1257+12 pulsar from orbital dynamics of its planets}
\shortauthors{L. Iorio}

\author{Lorenzo Iorio\altaffilmark{1} }
\affil{Ministero dell'Istruzione, dell'Universit$\grave{{\rm a}}$ e della Ricerca (M.I.U.R.). Fellow of the Royal Astronomical Society (F.R.A.S.). Permanent address: Viale Unit\`{a} di Italia 68, 70125, Bari (BA), Italy.}

\email{\emaila}

\begin{abstract}
We analytically compute the effects that a \textcolor{black}{pulsar's} mass variation, \textcolor{black}{whatever its physical origin may be, has} on the \textcolor{black}{standard Keplerian} changes $\Delta\tau_{\textcolor{black}{\rm Kep}}$ in the times of arrival of its pulses due to \textcolor{black}{potential} test particle companions, and on their orbital dynamics \textcolor{black}{over long time scales}. We apply our results to the planetary system of the PSR B1257+12 pulsar, located in the Galaxy at \textcolor{black}{$\sim 600$} pc from us, to phenomenologically constrain a putative accretion of non-annihilating dark matter on the hosting neutron star. By comparing our prediction for $\Delta\tau_{\dot M/M}$ to the root-mean-square accuracy of the timing residuals $\delta(\Delta\tau)=3.0\ \mu$s we find \textcolor{black}{for the mass variation rate} $\dot M/M\leq 1.3\times 10^{-6}$ yr$^{-1}$. Actually, considerations related to the pulsar's lifetime,  of the order of $\Delta t\sim 0.8$ Gyr, and to the currently accepted picture of the formation of its planets point toward a \textcolor{black}{tighter constrain on the} mass accretion rate, \textcolor{black}{i.e. $\dot M/M\leq 10^{-9}$ yr$^{-1}$.} Otherwise,  the planets would have formed at about $300-700$ au from PSR B1257+12, i.e. too far  with respect to the expected extension of $1-2$ au of the part of the protoplanetary disk containing the solid constituents from which they likely originated. \textcolor{black}{In fact, an even smaller upper limit,  $\dot M/M\leq 10^{-11}$ yr$^{-1}$, would  likely be more realistic to avoid certain technical inconsistencies with the quality of the fit of the timing data, performed by keeping the standard value $M=1.4 M_{\odot}$ fixed for the neutron star's mass.} Anyway, the entire pulsar data set should be re-processed \textcolor{black}{by explicitly modeling the mass variation rate and solving for it}. \textcolor{black}{Model-dependent} theoretical predictions for the pulsar's mass accretion, in the framework of the mirror matter scenario, yield a \textcolor{black}{mass} increment rate of about $10^{-16}$ yr$^{-1}$ for a value of the density of mirror matter $\rho_{\rm dm}$  as large as $10^{-17}$ g cm$^{-3}=5.6\times 10^6$ GeV cm$^{-3}$. Such a rate corresponds to a fractional mass variation of $\Delta M/M\sim 10^{-7}$ over the pulsar's lifetime. It would imply a formation of a black hole from the accreted dark matter inner core for values of the dark matter particle's mass $m_{\rm dm}$ larger than $3\times 10^3$ Gev, which are, thus, excluded since PSR B1257+12 is actually not such a kind of compact object. Instead, by assuming $\rho_{\rm dm}\sim 10^{-24}$ g cm$^{-3}=0.56$ GeV cm$^{-3}$, the mass accretion rate would be $\dot M/M\sim 10^{-23}$ yr$^{-1}$, with a fractional mass variation of the order of $\Delta M/M\sim 10^{-14}$. It rules out $m_{\rm dm}\geq 8\times 10^6$ Gev. \textcolor{black}{Extreme values  $\rho_{\rm dm}=1.8\times 10^{-13}$ g cm$^{-3}=10^{11}$ GeV cm$^{-3}$ for non-annihilating dark matter in central spike may \textcolor{black}{yield} the  constrain\textcolor{black}{t} $\dot M/M\leq 10^{-11}$ yr$^{-1}$; over $\Delta t=0.8$ Gyr, it rules out $m_{\rm dm}\geq 12$ Gev.}
 \end{abstract}

\section{Introduction}
\subsection{Need for Dark Matter}
An increasing number of observations at galactic, extragalactic and cosmological scales, if interpreted in the framework of the presently accepted Newtonian/Einsteinian laws of gravitation, requires the existence of huge amounts of a peculiar kind of matter  which does not emit electromagnetic radiation: the so-called Dark Matter (DM). It cannot be of baryonic nature. Indeed,
measurements of the baryon density in the Universe
using the Cosmic Microwave Background (CMB) spectrum and primordial nucleosynthesis constrain
the baryon density to a value less than $5\%$ of the critical density $\rho_{\rm crit}$. Instead, the total density of clustered matter, obtained from Supernov{\ae}-based
measurements of the recent expansion history of the Universe, CMB
measurements of the degree of spatial flatness, and measurements of the amount of matter
in galaxy structures obtained through big galaxy redshift surveys, is about $27\%$ of the critical density. Thus, about $22\%$ of it must exist in an exotic, unknown form. For reviews of both theoretical and observational aspects of the DM paradigm, see, e.g., \citet{Berg00,Gondolo,Berto05}.
\subsection{Dark Matter accretion onto astrophysical objects}
If DM exists, it should be present in all astrophysical objects; it  may both be there since their formation process and it may  subsequently be accreted from the surrounding environment.
In recent years much efforts have been devoted to investigate the phenomenon of possible capture of DM by neutron stars \citep{Gold89,Gou90,Bert08,Kouv08,Ciar09,Ciar010,deLav010,Kouv010,Gonza}. Indeed, such compact objects should efficiently capture DM because of their high matter density. The content of DM depend on the nature of its particles, the type of hosting celestial bodies and their history. Moreover,  new precise results from observations of neutron stars
are becoming more frequently available. Thus, at least in principle, they are considered as potentially useful tools to independently constraining various aspects of DM models like density, cross section and mass of their particles. Such parameters are also crucial in determining the capabilities of several Earth-based laboratory experiments like   CDMSI \citep{CDMSI}, CDMSII \citep{CDMSII}, DAMA/NaI \citep{dama1} and its successor DAMA/LIBRA \citep{Dama},
XENON10 \citep{XENON} and ZEPLIN III \citep{zeplin} aimed to directly detect DM.
\subsubsection{Self-annihilating Dark Matter and some consequences of its accretion on astrophysical objects}\lb{skazza}
According to the widely popular Weakly Interacting Massive Particle (WIMP) scenario, DM annihilates with itself and interacts
with the rest of the Standard Model (SM) via the weak interaction. The WIMP is typically defined as a stable, electrically
neutral, massive particle which arises naturally in supersymmetric SM extensions  \citep{Haber85}. A pair of WIMPs can annihilate, producing
ordinary particles and gamma rays. Self-annihilating particles captured by neutron stars would contribute to alter their outward appearances because the energy released in their annihilation would affect their internal and surface temperatures \citep{Gold89,Kouv08,deLav010,Kouv010,Gonza}. \textcolor{black}{On the other hand, WIMPs do not steadily accrete onto neutron stars and do not substantially  modify their inner structure by, e.g., inducing a massive DM core which may notably alter the gravitational collapse (see, instead, Section \ref{tragedy}).}.
\subsection{Non-annihilating Dark Matter. Mirror matter}
On the other hand,  models of DM exist in which it does not undergo self-annihilation \citep{Nussi,Kap92,Hoop05}; it may happen, for example, if DM is made of fermions, without the corresponding antifermions, or if DM consists of bosons and carries one sign of an additive conserved quantum number, but not the opposite sign. Among such scenarios there is the
mirror matter one \citep{Bli82,Khlo91,Khlo,Foot04,Okun07,Foot08,Blin010}. The possible existence of such an exotic form of matter was envisaged for the first time in the pioneeristic works by \citet{Lee} and, later, by \citet{Pom} and \citet{Pavs}; the modern form of such an idea was laid out  by \citet{Foot91}. Mirror matter arises if instead of (or in addition to) assuming
a symmetry between bosons and fermions, i.e. supersymmetry,
one assumes that nature is parity symmetric. In such a framework, in order to restore the parity symmetry violated by the weak interactions, the number of
particles in the Standard Model is doubled in such a way that the Universe is divided
into two sectors with opposite handedness that interact
mainly by gravity. On the other hand, parity can also be spontaneously broken depending on the Higgs potential \citep{rot1,rot2}.  While in the case of unbroken parity symmetry the masses of particles are the same as their mirror partners, in case of broken parity symmetry the mirror partners are lighter or heavier.
In regard to the interaction mechanisms among such putative mirror particles, the forces among them are mediated by mirror bosons. Now, with the exception of the graviton, none of the known bosons can be identical to their mirror partners. Mirror matter can interact with ordinary matter in a non-gravitational way only through the  so-called kinetic mixing of mirror bosons with ordinary bosons or via the exchange of
Holdom particles \citep{Holdom1,Holdom2}.  These interactions can only be very weak. That is  why mirror particles have  been suggested as DM candidates \citep{Bli82,Bli83,Kol,Khlo91,Hod}.
Putting effectively constraints on the masses of such kind of  stable  particles in Earth-based accelerator experiments is much more difficult than for self-annihilating candidates; in experiments like DAMA/NaI \citep{dama1} mirror DM would interact with ordinary matter via renormalizable photon-mirror
photon kinetic mixing, leading to a recoil energy-dependent cross section. Actually, mirror matter is one of the few DM candidates which can explain the positive DAMA/NaI \citep{dama1} dark matter signal whilst still being consistent with the null results of other DM experiments \citep{foo1,foo2}.
Another interesting feature of mirror matter is that it may overcome the difficulties that WIMP-based DM models have in explaining the opposite behaviors of DM in some colliding clusters of galaxies \citep{Zura09}. Indeed, while the behavior of Abell 520 points towards  a significant
self-interaction cross-section for DM \citep{abell}, the Bullet cluster (1E 0657-56), instead, behaves as a collisionless system \citep{bulletto}.  As a result, the inferred estimate on the DM
self-interaction cross section is well above the upper limit derived for
the Bullet cluster and exceeds by many orders the cross section magnitude
expected for WIMPs. On the contrary, mirror DM
models exhibit a greater flexibility; for them diverse behavior
of DM is a natural expectation \citep{Zura09}. On the other hand, \citet{Blin010} showed that the constraints on self-interaction cross-sections derived from observations of colliding clusters of galaxies are not real limits for individual particles if they form macroscopic bodies.
\subsubsection{Accretion of non-annihilating Dark Matter on astrophysical objects and some consequences of it}\lb{tragedy}
Some consequences of accretion of non-annihilating DM onto neutron stars have more or less recently been  investigated \citep{Bli83,Bert08,Ciar09,Ciar010,deLav010}. Basically, in this case a bulky mass of  DM would accumulate inside a neutron star without self-annihilating; \textcolor{black}{such kind of DM particles do not heat the star (cfr. with Section \ref{skazza})}.
The presence of such a DM core inside a neutron star may alter the usual mass-radius relation for such kind of astrophysical compact objects with potentially detectable consequences \citep{Ciar010}. Such an effect has been  calculated in the framework of mirror DM, but it is
qualitatively valid also for other kinds of non-annihilating DM that could form
stable cores inside neutron stars. Moreover, if the mass accretion continues steadily, the growing inner DM bulk may reach its own Chandrasekhar mass limit and collapse into a black hole, thus swallowing the hosting neutron star as well \citep{deLav010}. It must be noted that, in general, the Chandrasekhar \virg{dark} mass $M_{\rm Ch}^{(\rm dm)}$ would be smaller than in the usual case. Indeed, in terms of the Planck mass $M_{\rm Pl}$ \citep{deLav010},
\eqi M_{\rm Ch}^{(\rm dm)}\sim \rp{M_{\rm Pl}^3}{m^2_{\rm dm}},\eqf where $m_{\rm dm}$ is the mass of a DM particle which is larger than the ordinary nucleon mass $m_{\rm \textcolor{black}{nuc}}\sim 0.9$ Gev in most models. Such a dramatic outcome, $per\ se$ interesting, could, in principle, account for
the unexplained gamma ray bursts observed in the Universe
instead of resorting to the usual picture involving the coalescence of a neutron star with
another compact object.
%
%
%
%
\subsection{Overview of the paper}
In this paper we will consider other consequences of the accretion of non-annihilating DM by resorting to a specific scenario, i.e. the planetary system \citep{Wol92,rasio}
hosted by the PSR B1257+12 pulsar \citep{Wol90} situated in the Galaxy at high Galactic latitude at less than 1 kpc from us. In particular, we will look at some effects connected with the planets' orbital dynamics induced by a putative increment of non-annihilating DM experienced by the pulsar to derive bounds on such a phenomenon which are, then, used to constrain the resulting  dark core. A somewhat analogous study was performed with stellar motions around the Galactic Center to constrain DM annihilation proposed to explain the TeV gamma rays emanating from such a region of the Milky Way \citep{Hall06}. Concerning mirror matter and exoplanets,  \citet{exoplanets1,exoplanets2} suggested that the several close-in planetary companions of main-sequence stars discovered so far may be made up primarily of such a form of DM. \citet{planets} looked for mirror planets in our solar system itself.
Our approach can, in principle, be extended also to other similar scenarios involving a pulsar harboring compact or standard companion(s).
In Section \ref{calcolo} we will put phenomenological, model-independent constraints on $\dot M/M$ from the analysis of the impact that it may have on the \textcolor{black}{standard Keplerian variation of the} times of arrival  (TOAs) of pulsar's pulses \textcolor{black}{(Section \ref{kepl})} by comparing the present-day accuracy in their root-mean-square residuals with the analytically computed variation of TOAs due to $\dot M/M$ \textcolor{black}{(Section \ref{nonkep})}.  \textcolor{black}{We will also use some considerations on the age of the pulsar's system (Section \ref{nonkep2})}. Then, we will consider the global reduction of the spatial extension of the orbits of the planets of PSR B1257+12 due to its supposed mass accretion by contrasting our predictions of such an orbital shrinkage with the currently accepted picture of the birth of the system \textcolor{black}{(Section \ref{nonkep3})}. In Section \ref{moo} we will, \textcolor{black}{first}, compare our constraints to some recent predictions for the pulsar's mass accretion in terms of the mirror DM scenario \textcolor{black}{(Section \ref{pred1})}. \textcolor{black}{ Then, we will trace some consequences in terms of dark core collapse \textcolor{black}{(Section \ref{pred2})}}. Section \ref{conclusioni} is devoted to the conclusions.


\section{Planetary orbital effects of the mass accretion onto PSR B1257+12 }\lb{calcolo}
An ideal laboratory to study  certain consequences  of a putative accretion of non-annihilating DM on a neutron star in a purely phenomenological and model-independent way is represented, in principle, by a system hosting a pulsar orbited by one or more companions of planetary size; PSR B1257+12 and its three small Earth-sized planets represent one of such scenarios at\footnote{\textcolor{black}{Such an estimate for the distance is based on the Galactic electron distribution model by \citet{elec}. See also http://vizier.u-strasbg.fr/viz-bin/VizieR-S?PSR$\%$20B1257$\%$2b12.}}  \textcolor{black}{$600\pm 100$} pc \textcolor{black}{\citep{Kon03}} from us, with Galactic longitude and latitude  $l=311$ deg, $b=75$ deg, respectively. Indeed, in such  cases one can look at various effects induced by the pulsar's mass variation on the orbital motions of its planets.
Actually, the strategy devised below applies to any kind of putative mass variations of the primary.
\subsection{The Keplerian change in the times of arrival of the pulsar}\lb{kepl}
\textcolor{black}{The} direct observable is
the change $\Delta \tau$ in the pulsar's TOAs due to its orbiting partners.
The usual periodic variation $\Delta \tau_{\textcolor{black}{\rm Kep}}$ of TOAs resulting from the purely Keplerian motion of a pulsar around the center of mass of the star-planet system is \citep{Kon00}
\eqi \Delta \tau_{\textcolor{black}{\rm Kep}} = x\left[(\cos E -e)\sin\omega+\sqrt{1-e^2}\sin E\cos\omega\right],\lb{toa}\eqf
where $x\doteq a\sin I/c$, $[x]=$ T, is the projection of the  pulsar's semimajor axis $a$ with respect to the center of mass, $e$ is the eccentricity of the orbit, $E$ is the eccentric anomaly and $\omega$ is the argument of periastron.
Here $c$ denotes the speed of light in vacuum, $[c]=$ L T$^{-1}$. The inclination $I$ is the angle between the orbital angular momentum and the line-of-sight from the pulsar to us. The semi-major axis $a$ characterizes the size of a Keplerian ellipse: dimensionally,  $[a]=$ L. The eccentricity $e$ is an adimensional parameter which fixes the shape of a Keplerian ellipse. It is a non-negative real number which can assume all  values within $0\leq e <1$, where $e=0$ corresponds to a circle. The eccentric anomaly $E$ can be regarded as  a parametrization of the polar angle in the orbital plane. The longitude of periastron $\omega$ is an angle in the orbital plane which determines the position of the point of closest approach,  generally dubbed periapsis, with respect to a reference direction which is customarily assumed coincident with the line of the nodes. \textcolor{black}{The line of the nodes} is the intersection between the orbital plane and the plane of the sky, chosen in this case as reference plane. See, e.g., \citep{Roy} for basic concepts on orbital mechanics.

For a better comprehension, let us point out that \rfr{toa} yields the instantaneous value of $\Delta\tau_{\textcolor{black}{\rm Kep}}$, corresponding to a given value of $E$: indeed, there is a mapping between the eccentric anomaly and  time given by
 \eqi E=\mathcal{M}+\sum^{\infty}_{\ell=1}\left(\rp{2}{\ell}\right)J_{\ell}(\ell e)\sin(\ell \mathcal{M}),\ \ \mathcal{M}\doteq n(t-t_p),\lb{palla}\eqf
 where $J_{\ell}(\ell e),\ell=1,2,\ldots$ are the Bessel functions of first kind \citep{Bessel}, $\mathcal{M}$ is the mean anomaly \citep{Roy}, which is a parametrization of time, and $t_p$ is the time of passage at periastron. The series of \rfr{palla}  converges for all values of $e<1$ like a geometric series of ratio $\left(e\exp\sqrt{1-e^2}\right)/\left(1+\sqrt{1-e^2}\right)$ \citep{Wint,Ferna}.
 During an orbital revolution $E$ spans an angular interval of $2\pi$, in such a way that the timing $\tau$ does not remain constant, as it  happened if the pulsar was not perturbed by its companions, but exhibits a time-dependent, harmonic variation $\Delta\tau(E)$ which reveals the existence of other bodies in the system.
When a pulsar has $N$ companions, the TOA variations become \citep{Kon00}
\eqi \Delta \tau_{\textcolor{black}{\rm Kep}} = \sum_{j=1}^N = x_j\left[(\cos E_j -e_j)\sin\omega_j+\sqrt{1-e_j^2}\sin E_j\cos\omega_j\right].\eqf
\subsection{The non-Keplerian perturbations of the change in the times of arrival of the pulsar due to its mass variation}\lb{nonkep}
The further change in TOAs with respect to the purely Keplerian one of \rfr{toa} due to \textcolor{black}{a generic orbital perturbation caused by a small force which deviates from the largest Newtonian two-body, pointlike monopole acceleration $GM/r^2$ causing the well-known Keplerian motion} can  straightforwardly be worked out from \rfr{toa} itself by differentiating it with respect to the orbital parameters which undergo slow time-variations induced by \textcolor{black}{the perturbation considered. In the case of the PSR B1257+12 system it has recently been shown \citep{JOAA} that standard non-Keplerian  dynamical effects (departures from sphericity of the pulsar, 1PN, \textcolor{black}{Schwarzschild-like} corrections of order ${\mathcal{O}}(c^{-2})$)  which may cause departures from the main Keplerian picture are negligible, given the present-day accuracy in measuring some orbital characteristics like the orbital periods $P_{\rm b}$ of the planets.} \textcolor{black}{To better understand such points it is useful to note that the ratios of the pulsar's radius ($R\sim 10$ km) to the planetary orbital separations (about $0.2-0.5$ au, see Table \ref{tavolaparam}) are of the order of just $3.5-1.5\times 10^{-7}$. Thus, the point mass approximation is well justified, and the corrections induced by general relativity\footnote{\textcolor{black}{Since neutron stars are compact objects with strong
internal gravity, their gravitational fields must be fully described
within the framework of general relativity by using appropriate equations of state for their ultradense matter.} } to the values of both the mass and current  multipole moments \citep{Larak} of the external gravitational field of the neutron star are completely negligible as far as so distant orbital motions are concerned \citep{sibga,orbiz}.  The TOAs are, in principle, also affected  by effects concerning the propagation of the pulsar's electromagnetic waves in the distorted space-time (gravitational red-shift, Shapiro delay) \citep{Sta} usually accounted for by the post-Keplerian parameters\footnote{\textcolor{black}{The parameter $\gamma$ is the amplitude of the combined effect of the special relativistic time dilation and the gravitational red-shift, while $r$ is the amplitude of the Shapiro propagation delay caused by the gravitational field of the pulsar's companion. For an insightful qualitative description of such phenomena, which helps in understanding why they are negligible in the PSR B$1257+12$ system, see \citet{Kra}.}} $\gamma, r$ \citep{Dam}. However, inserting the figures of Table \ref{tavolaparam} into the analytical expressions for $\gamma$ and $r$ shows that they are completely negligible in the present case with respect to the accuracy in measuring the TOAs (see \rfr{resid} later). Thus, we will not further consider them in the following.}
\begin{table}
\centering
\caption{Relevant orbital parameters of the three planets A,B,C, of PSR B1257+12, from Table 2 of \citet{Kon03}. Here $a_{\rm P}$ are the planet semimajor axes. Figures in parentheses are the formal $1-\sigma$ uncertainties in the last digits quoted. The standard value $M=1.4M_{\odot}$ for the pulsar's mass has been \textcolor{black}{kept fixed} in deriving them, \textcolor{black}{i.e. it has not been included in the list of the  parameters to be solved-for in the fit of the pulsar's timing data}.}
\label{tavolaparam}
\begin{tabular}{llll}\hline
& A & B & C \\
\hline
$x$ (ms) & $0.0030(1)$ & $1.3106 (1)$ & $1.4134 (2)$\\
$a_{\rm P}$ (au) & $0.19$ & $0.36$ & $0.46$\\
$e$ & $0.0$ & $0.0186 (2)$ & $0.0252 (2)$ \\
$P_{\rm b}$ (d) & $25.262 (3)$ & $66.5419 (1)$ & $98.2114 (2)$ \\
$\omega$ (deg) & $0.0$ & $250.4 (6)$ & $108.3 (5)$ \\
\hline
\end{tabular}
\end{table}

\textcolor{black}{In the specific case of a putative pulsar's mass variation\footnote{\textcolor{black}{As recently shown both analytically and numerically by \citet{IorSRX}, such an effect is non-Keplerian in the sense that the resulting orbital motion of a test particle about the mass-varying primary is not a closed Keplerian ellipse. Note that this is a classical \textcolor{black}{orbital} effect, being the general relativistic \textcolor{black}{orbital} one totally negligible \citep{IorSRX}.}} with percent rate $\dot M/M$ the \textcolor{black}{perturbation to $\Delta\tau_{\rm Kep}$} }  is
\eqi \Delta \tau_{\dot M/M}= \left(\derp{\Delta \tau_{\textcolor{black}{\rm Kep}}}x\right) \Delta x + \left(\derp{\Delta \tau_{\textcolor{black}{\rm Kep}}}e \right)\Delta e + \left(\derp{\Delta \tau_{\textcolor{black}{\rm Kep}}}E \right)\Delta E + \left(\derp{\Delta \tau_{\textcolor{black}{\rm Kep}}}\omega\right)\Delta\omega,\lb{vaz}\eqf
with
\eqi
\left\{
\begin{array}{lll}
\derp{\Delta \tau_{\textcolor{black}{\rm Kep}}}x & = & \sqrt{1-e^2}\sin E\cos\omega+(\cos E-e)\sin\omega,\\ \\
\derp{\Delta \tau_{\textcolor{black}{\rm Kep}}}e & = & -x\left(\rp{e\sin E\cos\omega}{\sqrt{1-e^2}}+\sin\omega\right),\\ \\
\derp{\Delta \tau_{\textcolor{black}{\rm Kep}}}E & = & x\left(\sqrt{1-e^2}\cos E\cos\omega - \sin E\sin\omega\right),\\ \\
\derp{\Delta \tau_{\textcolor{black}{\rm Kep}}}\omega & = & x\left[(\cos E-e)\cos\omega-\sqrt{1-e^2}\sin E\sin\omega\right].
\end{array}
\right.
\eqf
In \rfr{vaz} the variations of $a,e,E,\omega$ due to $\dot M/M$ appear; $I$ is left unaffected by the mass variation of the primary \citep{IorSRX}; thus, $\Delta x=\Delta a\sin I/c$. They have been explicitly computed as functions of $E$ by \citet{IorSRX}:
\eqi
\left\{
\begin{array}{lll}
  \Delta a & = & -\left(\rp{\dot M}{M}\right)\rp{2ae}{n}\left(\rp{\sin E-E\cos E}{1-e\cos E}\right), \\ \\
  \Delta e  & = & -\left(\rp{\dot M}{M}\right)\rp{(1-e^2)}{n}\left(\rp{\sin E-E\cos E}{1-e\cos E}\right), \\ \\
  \Delta E & = & \left(\rp{\dot M}{M}\right)\rp{1}{n}\left[\mathcal{A}(E)+\mathcal{B}(E)+\mathcal{C}(E)\right], \\ \\
  \Delta\omega & = & -\left(\rp{\dot M}{M}\right)\rp{\sqrt{1-e^2}}{n e}\left[\rp{(1+e)(1-\cos E)-E\sin E}{1-e\cos E}\right],
\end{array}\lb{variz}
\right.
\eqf
where the coefficients $\mathcal{A},\mathcal{B},\mathcal{C}$ of the variation of the eccentric anomaly are
\eqi
\left\{
\begin{array}{lll}
\mathcal{A}(E) &=& \rp{E^2+2e(\cos E -1)}{1-e\cos E},\\ \\
\mathcal{B}(E) &=& \left(\rp{1-e^2}{e}\right)\rp{[(1+e)(1-\cos E)-E\sin E]}{(1-e\cos E)^2},\\ \\
\mathcal{C}(E) &=& -\rp{(1-e^2)\sin E(\sin E-e\cos E)}{(1-e\cos E)^2}.
\end{array}\lb{covariz}
\right.
\eqf
In \rfr{variz} $n\doteq 2\pi/P_{\rm b}$ is the Keplerian mean motion and $P_{\rm b}$ is the Keplerian orbital period.
For those readers not specifically acquainted with the methods of celestial mechanics it may be useful to point out some features of \rfr{variz}.
They represent the changes in $a,e,E,\omega$ induced by the considered perturbation at a given instant, to which correspond a given value of the eccentric anomaly $E$. Thus, they are not to be intended as the variations per orbit. Thus, \rfr{vaz} represents the instantaneous change in TOA, so that its variation per orbit must be evaluated by taking the difference between $\Delta\tau_{\dot M/M}$ computed at $E=2\pi$ and $\Delta\tau_{\dot M/M}$ computed at $E=0$. Incidentally, note also that the right-hand-sides of \rfr{variz} have correctly the same dimensions of the left-hand-sides. Indeed, $[n^{-1}]={\rm T}$, while $[\dot M/M]={\rm T}^{-1}$.
After $k=1,2,3,\dots$ revolutions, the TOA variation $\Delta^{(k)}_{\dot M/M}$ due to $\dot M/M$
  is
\eqi \Delta^{(k)}_{\dot M/M}\doteq \Delta \tau_{\dot M/M}(2k\pi)-\Delta \tau_{\dot M/M}(0)=k\left(\rp{\dot M}{M}\right)P_{\rm b}x\left[k2\pi\sqrt{\rp{1+e}{1-e}}\cos\omega -(1-e)\sin\omega\right].\lb{dtdm}\eqf

 As stated before, a useful application of the previously obtained results is represented by the planetary system of the  $6.2$ ms PSR B1257+12 pulsar.
Such a neutron star was discovered in 1990 during a high
Galactic latitude search for millisecond pulsars with the Arecibo radiotelescope
at 430 Hz \citep{Wol90}. Two years later, PSR B1257+12 turned out to be
orbited by at least two Earth-sized planets-dubbed B and C-along almost circular
paths \citep{Wol92}. In 1994 it was announced the discovery of a third, Moon-sized
planet-named A-in an inner, circular orbit \citep{Wol94}. Its presence, questioned by \citet{Sche97},
was subsequently confirmed in \citet{Kon99,Wol00}.

Concerning the accuracy in timing PSR B1257+12, covering 12 yr, a
detailed description of the data acquisition and the TOA
measurement process can be found in \citet{Wol00b}. The final
post-fit residuals for daily-averaged TOAs are characterized by a root-mean-square (rms) noise of\footnote{ \textcolor{black}{The relativistic contributions of order $\mathcal{O}(c^{-2})$ to $\Delta\tau$ due to gravitational time delay and the Shapiro delay are far smaller than \rfr{resid}.  } } \citep{Kon03}
 \eqi\delta(\Delta \tau)=3.0\ \mu {\rm s}.\lb{resid}\eqf

By equating \rfr{dtdm} for the three planets to  \rfr{resid} it is possible to obtain upper bounds for a putative mass variation experienced by PSR B1257+12; the results are in Table \ref{tavola}, \textcolor{black}{and do not depend on any specific model of non-annihilating DM}. Anyway, it should be pointed out that they should be considered just as preliminary, order-of-magnitude evaluations; actually, the entire pulsar timing data set should be re-processed by explicitly modeling the pulsar's mass variation as well, \textcolor{black}{and a dedicated solve-for parameter should be estimated in a least-square sense}.
\begin{table}
\centering
\caption{Order-of-magnitude upper bounds for the mass accretion rate of PSR  B1257+12 from \rfr{dtdm} applied to the three planets of the system. An accuracy of $\delta(\Delta\tau)=3.0$ $\mu$s in the TOAs residuals has been used \citep{Kon03}. A timing span of $\Delta t = 12$ yr has been assumed corresponding to approximately $k_{\rm A}\sim 170,k_{\rm B}\sim 66,k_{\rm C}\sim 45$ orbital revolutions for A,B,C. \textcolor{black}{They have been obtained in a phenomenological, model-independent way}.}
\label{tavola}
\begin{tabular}{llll}\hline
& A & B & C \\
\hline
$|\dot M/M|$ (yr$^{-1}$) & $7.6\times 10^{-5}$ & $1.3\times 10^{-6}$ & $1.9\times 10^{-6}$\\
\hline
\end{tabular}
\end{table}
From Table \ref{tavola} it can be noted that the tightest bound amounts to \eqi\left|\rp{\dot M}{M}\right|\leq 1.3\times 10^{-6}\ {\rm yr}^{-1}.\lb{vattinn}\eqf

\textcolor{black}{In Section \ref{nonkep2} and Section \ref{nonkep3}
we will independently check \rfr{vattinn} by inspecting its compatibility with other effects connected with the dynamical history of the PSR B1257+12 system.}
\subsection{Considerations from the age of the pulsar's planetary system}\lb{nonkep2}
Concerning the planets of PSR B1257+12,
it is just the case to review here that such a peculiar planetary system got formed \textcolor{black}{after} the death of the star progenitor of PSR B1257+12, not at its birth as a main sequence star, as in the usual planetary formation \citep{Pod}.
 \textcolor{black}{Thus, the} age of PSR B1257+12 as a neutron star cannot be smaller than $\sim 10^5$ yr, which is roughly the time required to form planets around millisecond pulsars from protoplanetary disks starting from the epoch of the birth of the pulsar itself \textcolor{black}{as stellar corpse} \citep{proto}; this implies that, by setting $\Delta t=10^5$ yr, \textcolor{black}{\rfr{vattinn} yields}
\eqi \rp{\Delta M}{M}\lesssim 0.13. \lb{troppo} \eqf
Actually, such a constrain seems to be too large. Indeed, apart from the fact that it would imply that, in the framework of the non-annihilating DM accretion scenario, PSR B1257+12 should have already become a black hole, \textcolor{black}{at least for certain values of DM particle's mass (see Figure 8 of \citet{deLav010} and \rfr{antobizz} below)}, on the other hand its mass should \textcolor{black}{now} be as large as $1.53M_{\odot}$ \textcolor{black}{with respect to an assumed standard value of $M=1.4 M_{\odot}$ at its birth}. \textcolor{black}{But} the post-fit residuals in the TOAs of PSR B1257+12  are statistically compatible with zero, \textcolor{black}{i.e. they do not exhibit anything anomalous at a statistically significant level. To this aim, it should be noted that such residuals have been obtained just for $M=1.4 M_{\odot}$.} Stated differently, if the mass of PSR B1257+12 was really larger than the standard value by roughly $13\%$, the timing residuals, constructed just by keeping fixed the pulsar's mass to the standard value, should have retained some statistically significant non-zero features, \textcolor{black}{which is not the case.} Thus, we conclude that $|\dot M/M|$ should be actually  smaller than $10^{-6}$ yr$^{-1}$ by likely one-two orders of magnitude. In fact, a more quantitatively precise  evaluation of it would require a re-processing of the pulsar timing data with different values of its mass to check the level at which departures from the standard value $M=1.4 M_{\odot}$ cease to affect the TOAs post-fit residuals. It must be noted that such considerations may be quite optimistic since the age of PSR B1257+12 may be as large as $\sim 1 $ Gyr \citep{eta}. The age of the pulsar
is estimated from its spin-down timescale $\mathcal{P}/2\dot {\mathcal{P}}$ \citep{eta} to be $\lesssim 1$ Gyr since the pulsar's spin period is $\mathcal{P}=6.2$ ms and its variation is $\dot{\mathcal{P}}=1.1 \times 10^{-16}$ ms s$^{-1}$ \citep{Kon03}. In this case, by repeating the previous calculations, $\dot M/M$ should be, \textcolor{black}{perhaps}, of the order of $10^{-11}$ yr$^{-1}$ or less to avoid striking contradictions with the lacking of likely statistically significant patterns in the TOAs residuals obtained for $M=1.4M_{\odot}$.
\textcolor{black}{All in all, it should be recalled that to be more quantitative the phenomenon of mass variation of the pulsar should be explicitly modeled in the dynamical force models used in the timing processor systems and explicitly solved-for in the consequent re-analysis of the timing data.}
\subsection{The effect of the pulsar's mass variation on the orbits of its planets throughout their history}\lb{nonkep3}
An independent check of the previous considerations can be obtained by inspecting another consequence of the mass variation of the primary on the orbital motions of its test particle companions: the change in the size of their orbits. Indeed, if PSR B1257+12 was really accreting its mass since its birth, the spatial extension of its planetary system should have been larger than now when it formed. Although no observationally determined quantities related to such a putative variation of its size are available for the PSR B1257+12 system, it is interesting to inspect the consequences that such an orbital shrinking may have on timescales comparable to the pulsar's lifetime and compare them to the currently accepted picture of the formation of its planetary system \citep{eta,proto}.
From the expression of the Keplerian planet's astrocentric distance \citep{Roy}
\eqi r=a(1-e\cos E),\eqf
it follows
\eqi \Delta r (E) = (1-e\cos E)\ \Delta a-a\cos E\ \Delta e + ae\sin E\ \Delta E.\lb{bafana}\eqf It
 represents the instantaneous departure of the astrocentric distance with respect to the unperturbed, Keplerian one.
Note that \rfr{bafana} agrees with the results obtained in, e.g., \citet{Cas93}.
In the case of a mass variation, from \rfr{variz}-\rfr{covariz} we have \citep{IorSRX}
\eqi \Delta r_{\dot M/M}(E) = \left(\rp{\dot M}{M}\right)\rp{a}{n}\left[\mathcal{D}(E)+\mathcal{F}(E)\right], \lb{mega}\eqf
with
\eqi
\left\{
\begin{array}{lll}
\mathcal{D}(E) & = & e\left[-2(\sin E-E \cos E) + \rp{\sin E\left[E^2 + 2e(\cos E -1)\right]}{1-e\cos E} -\rp{(1-e^2)\sin^2 E(\sin E-e\cos E)}{(1-e\cos E)^2} \right],\\\\
\mathcal{F}(E) & = & \left(\rp{1-e^2}{1-e\cos E}\right)\left\{
\cos E(\sin E - E \cos E) + \sin E\left[\rp{(1+e)(1-\cos E)-E \sin E}{1-e\cos E}\right]
\right\}.
\end{array}
\right.
\lb{mega2}\eqf
It turns out from \rfr{mega} and \rfr{mega2}  that the shift in the star-planet distance $\delta^{(k)}_{\dot M/M}$ after $k$ orbital revolutions is
\eqi \delta^{(k)}_{\dot M/M}
\doteq\Delta r_{\dot M/M} (2k\pi) - \Delta r_{\dot M/M} (0) = -k\left(\rp{\dot M}{M}\right)P_{\rm b}a(1-e).\lb{deltaerre}\eqf
As expected, \rfr{deltaerre} tells us that the orbit gets smaller for a mass increase of the primary, i.e. $ \delta^{(k)}_{\dot M/M}<0$ for $\dot M/M>0$; note that here $a$ denotes the relative star-planet semimajor axis whose values are listed in Table \ref{tavolaparam}.
The application of the upper bound of \rfr{vattinn} to \rfr{deltaerre} over a past time span of the order of the system's lifetime, i.e $\Delta t\sim 1$ Gyr, yields the implausible results listed in Table \ref{seedomain}: A,B,C, should have formed at hundreds au far from the pulsar to have reached nowadays their present astrocentric distances under the action of a mass accretion experienced by their primary as large as that of \rfr{vattinn}.
\begin{table}
\centering
\caption{Variations $\delta_{\dot M/M}^{(k)}$  of the astrocentric distances, in au,  of the three planets of PSR B1257+12 due to $\dot M/M=1.3\times 10^{-6}$ yr$^{-1}$. A time span of $\Delta t = -1$ Gyr has been assumed corresponding to approximately $k_{\rm A}\sim 1\times 10^{10},k_{\rm B}\sim 5\times 10^9,k_{\rm C}\sim 3\times 10^9$ orbital revolutions for A,B,C.}
\label{seedomain}
\begin{tabular}{llll}\hline
& A & B & C \\
\hline
$\delta_{\dot M/M}^{(k)}$ (au) & $-304$ & $-565$ & $-717$\\
\hline
\end{tabular}
\end{table}
A mass accretion rate three orders of magnitude smaller than \rfr{vattinn} for PSR B1257+12, i.e. of the order of $10^{-9}$ yr$^{-1}$, would, instead, yield primeval astrocentric distances as large as about 0.7 au, which is substantially in agreement with the maximum extension  of $\sim 1-2$ au of the part of the protoplanetary disk containing solid materials from which A,B,C likely originated \citep{proto}.
\textcolor{black}{On the other hand, an age of the pulsar as large as about 1 Gyr would yield $\Delta M/M\sim 0.8-1$, so that the same problems encountered in Section \ref{nonkep2} would occur. Also in this case, a mass accretion rate two orders of magnitude smaller, i.e.  $\dot M/M\sim 10^{-11}$ yr$^{-1}$ which would be well compatible with the distances at which the planets should have formed, would cure them.}
\section{Confrontation with some \textcolor{black}{theoretical scenarios}}\lb{moo}
\subsection{Predictions of some scenarios for mirror matter accretion}\lb{pred1}
Let us, now, consider the predictions for the accretion of non-annihilating DM. \textcolor{black}{It is just the case to briefly recall that while WIMPs may heat a neutron star without creating a particularly massive dark core inside, on the contrary, mirror particles do not heat a neutron star and may form a dark core with substantial mass that modifies the structure of the star.}
Reasoning in terms of mirror matter, its distribution in galaxies  is expected to be non-homogeneous since it should form complex structures similarly to as ordinary baryons do. Thus, the accretion rate of mirror matter by neutron stars should depend on the location and history of each star. Of course, also the structure of the hidden mirror sector, which is unknown, is relevant.
\citet{Ciar09} yield
\eqi \dot M = 10^7\ {\rm kg\ s}^{-1}\lb{limitez}\eqf
as possible upper bound of accretion of mirror matter onto a neutron star.
Such an estimate is based on
\citep{Shap83}
\eqi \dot M=\left(\rp{\rho_{\rm dm}}{10^{-24}\ {\rm g\ cm}^{-3}}\right)\left(\rp{10\ {\rm km\ s}^{-1}}{v_{\rm dm}}\right)\left(\rp{M}{M_{\odot}}\right)^2\ \textcolor{black}{\rm kg\ s^{-1}}\lb{kazza},\eqf
which has been derived in the hypothesis that the distribution of mirror particles is isotropic and monoenergetic at large distance from the neutron star.
In obtaining \rfr{limitez} from \rfr{kazza} it has been assumed that the density of the interstellar medium in the mirror sector may be up to one order of magnitude larger than the value $\rho_{\rm dm}=10^{-18}$ g cm$^{-3}=5.6\times 10^5$ Gev cm$^{-3}$ of the ordinary baryonic giant molecular clouds \citep{Ciar09}, who tacitly use $v_{\rm dm}=10$ km s$^{-1}$ and $M=M_{\odot}$ as well. Note that such a figure is orders of magnitude larger than the currently accepted value of the dark halo density at about 10 kpc from the Galactic Center which is of the order of\footnote{\textcolor{black}{1 g cm$^{-3}$ corresponds to $5.6\times 10^{23}$ Gev cm$^{-3}$.}}  $\sim 10^{-24}$ g cm$^{-3}=0.56$ Gev cm$^{-3}$ \citep{Berg00,Berto05,deLav010}. \textcolor{black}{Incidentally, let us note from Figure 1 of \citet{deLav010} that all the Einasto DM density profiles substantially converge to such a figure for $\rho_{\rm dm}$ at Galactocentric distances larger than 10 kpc, so that the uncertainty of 100 pc \citep{Kon03} in the heliocentric distance of PSR B1257+12 is of no concern for us. }
\textcolor{black}{It is just the case of stressing that the estimate from \rfr{kazza} has to be intended as referred to a local quantity. Indeed, the average DM density in the mirror matter scenario is comparable to that of other DM models. Contrary to them, mirror matter is believed to have the possibility of forming structures having higher density than the average DM one. }
 In the case of PSR B1257+12, by assuming that it steadily accreted DM throughout its lifetime
\eqi\Delta t = {\mathcal{P}}/2\dot{\mathcal{P}} = 0.893\ {\rm Gyr},\lb{tempopsr}\eqf
\rfr{limitez} would translate into
a mass accretion rate of
\eqi \left.\rp{\dot M}{M}\right|_{\rm mirror}=1.1\times 10^{-16}\ {\rm yr}^{-1}.\lb{massavar}\eqf It is compatible with the phenomenological constraints previously obtained for the PSR B1257+12 system.
According to \rfr{massavar} and \rfr{tempopsr},  PSR B1257+12 would have accreted a mass fraction
\eqi \rp{\Delta M}{M}=1.0\times 10^{-7}\lb{pizza}\eqf
during its lifetime.
If, instead, we use \rfr{kazza} with $\rho_{\rm dm}\sim 10^{-24}$ g cm$^{-3}=0.56$ Gev cm$^{-3}$, it yields an accretion rate of approximately \eqi \left.\rp{\dot M}{M}\right|_{\rm mirror} = 2\times 10^{-23}\ {\rm yr}^{-1}\lb{kazza2}\eqf corresponding to a fractional mass variation of PSR B1257 +12 over its lifetime of
\eqi \rp{\Delta M}{M}=1.9\times 10^{-14}.\lb{pizza2}\eqf  Also in this case, such figures are compatible with the constraints phenomenologically obtained.
Let us mention that, recently, \citet{Gonza} yielded a mass accretion rate for pulsars in the solar neighborhood of the order of
\eqi \dot M\leq 9\times 10^{-25}\ M_{\odot}\ {\rm yr}^{-1},\eqf although it seems that they considered self-annihilating DM.
\textcolor{black}{
Let us note that that a mass accretion rate of about $10^{-11}$ yr$^{-1}$, which would be in agreement with both the planetary formation scenario and the fit of the timing data for PSR B1257+12 (Section \ref{nonkep2}-Section \ref{nonkep3}), could be obtained from \rfr{kazza} by using the extreme limit $\rho_{\rm dm}=1.8\times 10^{-13}$ g cm$^{-3}=10^{11}$ GeV cm$^{-3}$ of the predictions by \citet{BerMer05}
for non-annihilating DM in a central spike.
}

Another approach that, in principle, one could follow consists of performing specific numerical simulations coming up with a quantitative estimate of the probability that PSR B1257+12 eventually emerged where it is now located starting from some detailed evolutionary models of the distribution of the mirror matter in the Galaxy taking into account issues like its form, clumpiness, density of clumpiness, etc, and by using  different values of its density.
This is beyond the scope of the present paper. Let us  mention that such relatively \virg{fine-graining} simulations of mirror matter distribution at the level of the Milky Way sub-structure have not (yet?) been implemented; some work has been performed  at a more general level concerning galactic haloes \citep{Moha},  and  cosmological large scale structure formation \citep{igna,stru}.
\subsection{Consequences in terms of dark core collapse \textcolor{black}{for WIMPs only}}\lb{pred2}
Let us interpret our   results in terms of dark core collapse.
In such a framework, the limiting case of the mass variation of a neutron star occurs when its putative dark core has reached its Chandrasekhar mass, i.e.
\eqi \rp{\Delta M}{M}^{(\rm coll)}\doteq \rp{[M^{(\rm dm)}_{\rm Ch}+M]-M}{M}=\rp{M^{(\rm dm)}_{\rm Ch}}{M}\sim \rp{M^3_{\rm Pl}}{m_{\rm dm}^2M}=\rp{1.1655\ {\rm Gev^2}}{m^2_{\rm dm}}.\lb{dmm}\eqf
According to \textcolor{black}{the values for $m_{\rm dm}$ reported in} Table \ref{masse},
\begin{table}
\centering
\caption{Values, in Gev, of some mass energies used in the text. $M_{\odot}$ is the mass of the Sun, $M_{\rm Pl}$ is the Planck mass, $m_{\rm dm}^{(\rm min)}$, $m_{\rm dm}^{(\rm max)}$ are the minimum and maximum masses of a \textcolor{black}{non-annihilating} DM particle according to Figure \textcolor{black}{8} of \citet{deLav010}. \textcolor{black}{ Actually, in the mirror matter scenario  $m_{\rm dm}^{(\rm min)}$ can be much smaller than \textcolor{black}{50} Gev \citep{rot2}.} }
\label{masse}
\begin{tabular}{llll}
\hline
$M_{\odot}$ (Gev) & $M_{\rm Pl}$ (Gev)  & $m_{\rm dm}^{(\rm min)}$ (Gev) & $m_{\rm dm}^{(\rm max)}$ (Gev)\\
\hline
$1.1157\times 10^{57}$ & $1.22105\times 10^{19}$ & \textcolor{black}{50} & $10^{\textcolor{black}{8}}$ \\
\hline
\end{tabular}
\end{table}
the maximum and minimum values for \rfr{dmm} are, by assuming $M=1.4M_{\odot}$,
\eqi
\begin{array}{lll}
\left.\rp{\Delta M}{M}^{(\rm coll)}\right|_{\rm min} &=& 1.1\times 10^{-\textcolor{black}{16}}, \\ \\
\left.\rp{\Delta M}{M}^{(\rm coll)}\right|_{\rm max} &=& \textcolor{black}{4.6}\times 10^{-4}.
\end{array}\lb{antobizz}
\eqf
%
 Since, evidently, PSR B1257+12 has not (yet?) become a black hole,
 \textcolor{black}{it must be
 \eqi \left(\rp{\dot M}{M}\right)_{\rm \textcolor{black}{non-ann}}\Delta t < \rp{\Delta M}{M}^{(\rm coll)}.\lb{cord}\eqf
 According \textcolor{black}{to} our \textcolor{black}{model-dependent} estimates on $\Delta M/M$ of \rfr{pizza} and \rfr{pizza2}, the condition of \rfr{cord} is clearly satisfied for the smallest admissible values of $m_{\rm dm}$ connected to the maximum figure in \rfr{antobizz}. To this aim, let us incidentally note that in the mirror matter scenario the lower limit for $m_{\rm dm}$ can actually be much smaller than that quoted here in Table \ref{masse}; see, e.g., \citet{rot2}. On the contrary, \rfr{cord} does not hold for the largest supposed values of $m_{\rm dm}$ yielding the smallest values in \rfr{antobizz}. Thus, \rfr{cord} is able to put upper bounds on $m_{\rm dm}$.
 To this aim,} \rfr{massavar}  allow\textcolor{black}{s} to rule out DM candidates with masses larger than \eqi m_{\rm dm} = \sqrt{\rp{M^3_{\rm Pl}}{M \left(\dot M/M\right)_{\rm \textcolor{black}{non-ann}}\Delta t}}= 3\times 10^3\ {\rm Gev}\eqf because they would yield limiting core-collapse fractional mass variations smaller than  \rfr{pizza}.
Instead, the much smaller figure of \rfr{kazza2}  yields an upper bound on the DM particle's mass
\eqi m_{\rm dm} = 8\times 10^6\ {\rm Gev}.\eqf

\textcolor{black}{
Moving to the phenomenological, model-independent constraints on $\dot M/M$ of Section \ref{nonkep}- Section \ref{nonkep3}, it is apparent that they do not fulfil the condition of \rfr{cord}, at least for the values of $m_{\rm dm}$ considered in Figure 8 of \citet{deLav010}. However, as already noted, in the mirror matter scenario the mass of the DM particles can be much smaller than 50 Gev \citep{rot2}. A mass accretion rate of $10^{-9}$ yr$^{-1}$, which is compatible with the planetary formation history (Section \ref{nonkep3}), would yield over $\Delta t=0.8$ Gyr an upper bound on the DM particle's mass of $1.2$ GeV. If, instead, we take $\dot M/M = 10^{-11}$ yr$^{-1}$, which  would yield just a $0.8\%$ departure of the pulsar's mass from its standard value over $\Delta t=0.8$ Gyr (Section \ref{nonkep2}-Section \ref{nonkep3}), the upper bound is $m_{\rm dm}< 12$ Gev.
}
\section{Summary and conclusions}\lb{conclusioni}
 \textcolor{black}{A} neutron star in the Galaxy may accrete non-annihilating DM at a rate  up to about $\dot M=10^7$ kg s$^{-1}$ for $\rho_{\rm dm}\sim 10^{-17}$ g cm$^{-3}\sim 5.6\times 10^{6}$ GeV cm$^{-3}$ according to some authors; if, instead, we assume $\rho_{\rm dm}\sim 10^{-24}$ g cm$^{-3}\sim 0.56$ GeV cm$^{-3}$, the mass accretion rate would be about $\dot M =1$ kg s$^{-1}$; \textcolor{black}{extreme values like $\rho_{\rm dm}\sim 10^{-13}$ g cm$^{-3}\sim 10^{11}$ GeV cm$^{-3}$ are also possible according to other researchers, yielding a mass accretion rate of the order of $10^{-11}$ yr$^{-1}$ }. The steady accumulation of non-annihilating DM inside a neutron star may yield the formation of an inner dark core. If it reaches its own Chandrasekhar mass $M_{\rm Ch}^{(\rm dm)}\sim M_{\rm Pl}^3/m^2_{\rm dm}$, which should be \textcolor{black}{different from}  that made of ordinary baryons in view of the expected \textcolor{black}{different} mass $m_{\rm dm}$ of the the DM particles with respect to standard nucleons, such a core may collapse into a black hole, thus destroying the hosting neutron star. We used the PSR B1257+12 millisecond pulsar, located at \textcolor{black}{$\sim 600$} pc from us and hosting a planetary system of three Earth-sized companions, to put phenomenologically constraints on such a putative mass accretion rate $\dot M/M$ by looking at some dynamical orbital effects \textcolor{black}{affected by such a phenomenon. We also exploited the expected lifetime of the neutron star. }

In the case of the PSR B1257+12, the  aforementioned predicted mass variation rates would correspond, for $M=1.4M_{\odot}$, to $\dot M/M=\sim 10^{-16}$ yr$^{-1}$, $\dot M/M\sim 10^{-23}$ yr$^{-1}$,  \textcolor{black}{and $\dot M/M\sim 10^{-11}$ yr$^{-1}$}, respectively. They are \textcolor{black}{all} compatible with the upper bound $\dot M/M\leq 1.3\times 10^{-6}$ yr$^{-1}$  phenomenologically obtained by comparing the analytically calculated \textcolor{black}{perturbations by}  $\dot M/M$ on the \textcolor{black}{standard Keplerian} variations $\Delta\tau_{\textcolor{black}{\rm Kep}}$ of the times of arrival of the pulsar's pulses due to the presence of its planets to the root-mean-square residuals $\delta(\Delta\tau)=3.0\ \mu$s of the TOAs.
Smaller \textcolor{black}{phenomenological} constraints by some orders of magnitude \textcolor{black}{($\dot M/M\sim 10^{-9}$ yr$^{-1}$)} come from the confrontation of the predicted global shrinking of the orbits of the planets during the pulsar's lifetime $\Delta t\sim 0.8$ Gyr to the currently accepted picture of their formation from a protoplanetary disk whose part containing solid particles extended just for  $1-2$ au; for $\dot M/M\sim 10^{-6}$ yr$^{-1}$ the planets would have formed at approximately $300-700$ au from the pulsar. \textcolor{black}{Given the pulsar's lifetime, in order to avoid possible contradictions with the fact that the standard value of the pulsar's mass $M=1.4M_{\odot}$, kept fixed in its timing data fitting, did not destroy the goodness of the fit at a statistically significant level, a smaller mass variation rate $\dot M/M\leq 10^{-11}$ yr$^{-1}$ may be considered more plausible. However,  to be more quantitative, the entire pulsar timing data set should be re-processed by explicitly including the effect of $\dot M/M$  in the dynamical models used.}
According to \textcolor{black}{some}  predictions of the \textcolor{black}{non-annihilating DM} scenario, during its lifetime PSR B1257+12 should have accreted a mass fraction as large as $\Delta M/M=1.0\times 10^{-7}$ ($\rho_{\rm dm}\sim 10^{-17}$ g cm$^{-3}\sim 10^6$ GeV cm$^{-3}$), or $\Delta M/M=1.9\times 10^{-14}$ ($\rho_{\rm dm}\sim 10^{-24}$ g cm$^{-3}=0.56$ GeV cm$^{-3}$), respectively. Since it is not (yet?) a black hole, such figures exclude values for the DM particle's mass larger than $3\times 10^3$ Gev and $8\times 10^6$ Gev, respectively. \textcolor{black}{Instead, $\dot M/M\sim 10^{-11}$ yr$^{-1}$ and the pulsar's lifetime would imply a fractional mass accretion $\Delta M/M$ which rules out $m_{\rm dm}\geq 12$ GeV.}

The approach followed here may be, in principle, extended to other scenarios involving one pulsar hosting compact or planetary companions.

\section*{Acknowledgments}
I thank some anonymous referees for their valuable comments and remarks which greatly improved the manuscript.


\end{document}